\documentclass[a4paper, 10pt, onecolumn]{quantumarticle}
\pdfoutput=1

\usepackage{amsmath}
\usepackage{url}
\usepackage{braket}

\usepackage{array}   
\newcolumntype{L}{>{$}l<{$}} 

\usepackage{qcircuit}

\usepackage{fancyhdr}
\usepackage{fancyvrb}
\usepackage{xcolor}
\usepackage{hyperref}
\usepackage{natbib}

\usepackage{mathdots}

\usepackage{graphicx}
\setkeys{Gin}{width=\linewidth,totalheight=\textheight,keepaspectratio}
\graphicspath{{figures/}}

\usepackage{amssymb, amsthm, bm}

\newcommand{\be}{\begin{equation}}
\newcommand{\ee}{\end{equation}}
\newcommand{\bea}{\begin{eqnarray}}
\newcommand{\eea}{\end{eqnarray}}

\usepackage{graphicx}
\usepackage{dcolumn}
\usepackage{bm}
\usepackage{hyperref}
\usepackage{color}
\usepackage[caption=false]{subfig}
\usepackage{microtype}
\usepackage{verbatim} 
\usepackage{amsmath,amsfonts,amssymb,amsthm}
\usepackage[algoruled, vlined, lined ,boxed,commentsnumbered]{algorithm2e}
\usepackage{mathtools}
\usepackage{physics}

\usepackage{tikz}
\hypersetup{%
        colorlinks=true,
        linkcolor=blue,
        citecolor=blue,
        urlcolor=blue,
      }
\usepackage[euler]{textgreek}
\usepackage[ugly]{units}
\usepackage{xcolor}
\usepackage{ wasysym }

\graphicspath{figures/}

\theoremstyle{definition}

\begin{document}
\title{Hardware optimized parity check gates for superconducting surface codes}
\author{Matthew J. Reagor}
\email{matt@rigetti.com}
\affiliation{Rigetti Computing, Berkeley, CA 94710}
\author{Thomas C. Bohdanowicz}
\email{thom.bohdanowicz@gs.com}
\affiliation{Goldman, Sachs \& Co., New York, NY}
\author{David Rodr\'iguez P\'erez}
\affiliation{Rigetti Computing, Berkeley, CA 94710}
\author{Eyob A. Sete}
\affiliation{Rigetti Computing, Berkeley, CA 94710}
\author{William J. Zeng}
\affiliation{Goldman, Sachs \& Co., New York, NY}

\date{November 11, 2022}

\maketitle

\begin{abstract}
Error correcting codes use multi-qubit measurements to realize fault-tolerant quantum logic steps. In fact, the resources needed to scale-up fault-tolerant quantum computing hardware are largely set by this task. Tailoring next-generation processors for joint measurements, therefore, could result in improvements to speed, accuracy, or cost---accelerating the development large-scale quantum computers. Here, we motivate such explorations by analyzing an unconventional surface code based on multi-body interactions between superconducting transmon qubits. Our central consideration, Hardware Optimized Parity (HOP) gates, achieves stabilizer-type measurements through simultaneous multi-qubit conditional phase accumulation. Despite the multi-body effects that underpin this approach, our estimates of logical faults suggest that this design can be at least as robust to realistic noise as conventional designs. We show a higher threshold of $1.25 \times 10^{-3}$ compared to the standard code's $0.79 \times 10^{-3}$. However, in the HOP code the logical error rate decreases more slowly with decreasing physical error rate. Our results point to a fruitful path forward towards extending gate-model platforms for error correction at the dawn of its empirical development.
\end{abstract}

\section{Introduction} 
Surface codes~\citep{Bravyi1998,Dennis_2002,Fowler_2012} and related methods for stabilizer quantum error correction on 2D lattices~\citep{Bravyi2012,Terhal_2015,Li2019} are leading candidates for large-scale quantum computers based on solid-state platforms, such as superconducting qubits. As these types of processors demonstrate increasingly sophisticated error correction protocols~\citep{Kelly_2015,Andersen_2020,Marques_2021,Chen_2021,Krinner_2022}, the opportunity to tailor code design to superconducting hardware~\citep{Michael_2016,Tuckett_2018} and vice-versa~\citep{Mirrahimi_2014,Royer_2018,Chamberland_2022} promises to accelerate progress towards fault-tolerant algorithms. Indeed, specialisation has already played an important role in this respect---the first bit-flip correction mechanisms on superconductors leveraged a non-standard Toffoli gate~\citep{Reed_2012}, and achieving break-even error rates between physical and logical qubits for superconductors relied on an auxiliary memory to the transmon qubit~\citep{Ofek_2016}. Yet, hardware that has been purpose-made for error correction is not necessarily also optimal for studying near-term quantum algorithms\footnote{such as variational quantum-classical methods~\citep{Cerezo_2021}}. As a consequence, systems that have been scaled-up to the order of one-hundred qubits, to date, are exclusively gate-model devices based on planar superconducting transmon circuits.

\begin{figure}
\centering
\includegraphics[width=0.8\textwidth]{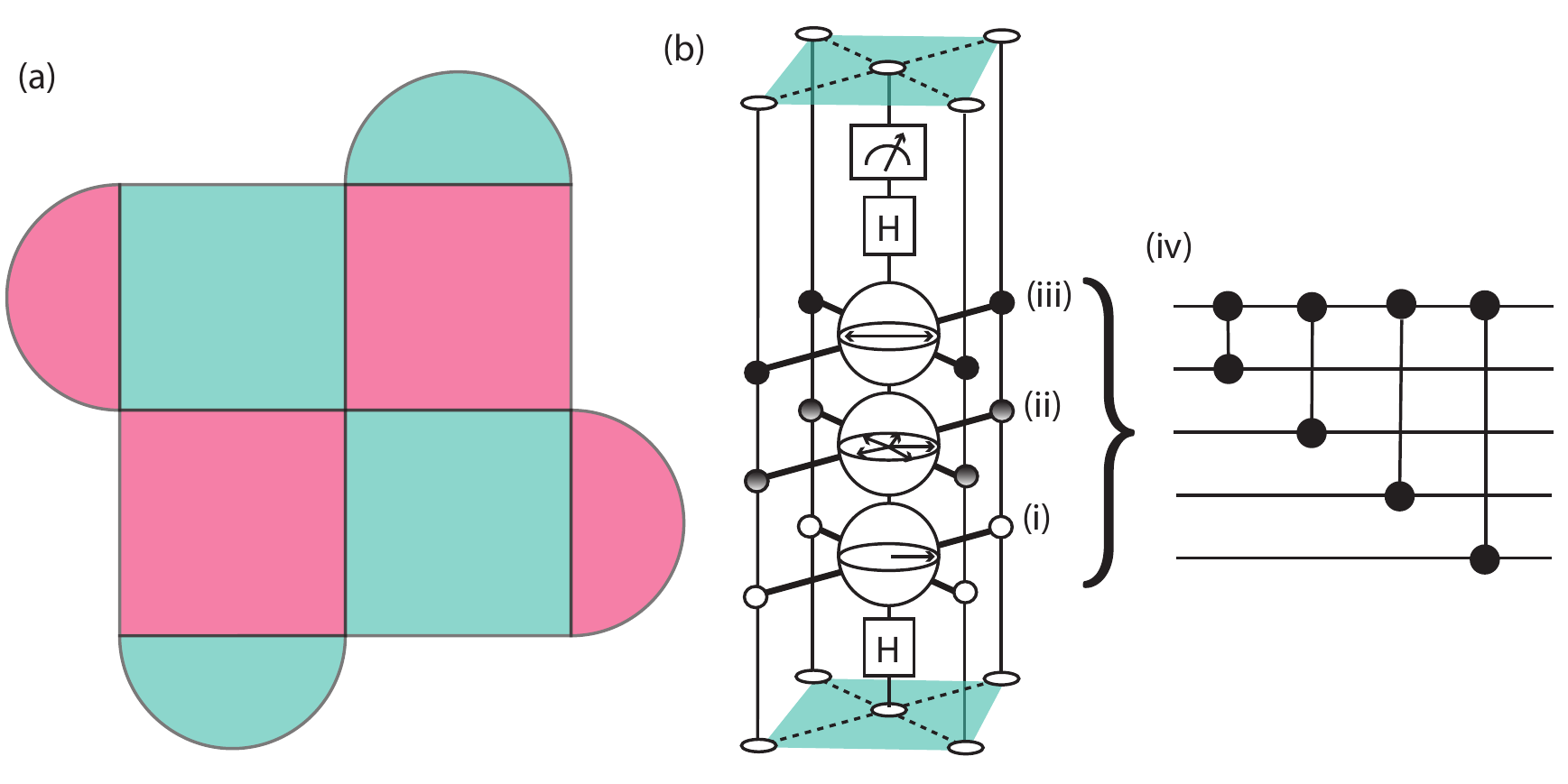}
\caption{\label{fig:hop_overview} Concept surface code design. (a) A small patch of rotated surface code with Z-checks shown in teal and X-checks shown in magenta. Data qubits are located at the vertices of the tessellation with a single stabilizer check qubit within each tile. (b) Performing a Z-check with the Hardware-Optimized Parity (HOP) gate. Dashed lines on this square indicate the presence of pairwise dispersive $ZZ$ interaction between the stabilizer check qubit, located at the center of the square, and the four data qubits, located at the vertices. The Z-check protocol begins with an unconditional Hadamard gate applied to the stabilizer check qubit preparing the qubit along the X-axis of its Bloch sphere. Activating the gate and evolving the five-qubit system under the interaction Hamiltonian accumulates the desired phase gate, wherein at (i) no entanglement is present, (ii) phase collapse, (iii) register parity mapped to the $\pm X$ states of the check qubit. The time evolution (i-iii) is equivalent to the four $CZ$ circuit shown in (iv). Finally, a second unconditional Hadamard gate rotates the check qubit to the measurement $Z$ basis. The sequence for $X$-checks is equivalent up to additional Hadamard gates on the data qubits before and after the sequence shown for $Z$-checks.}
\end{figure}

Motivated by progress extending the reach of these gate-model platforms with application-specific logic---including gates for chemistry~\citep{Foxen_2020}, many-body complex systems~\citep{Blok_2021}, and optimization~\citep{Lacroix_2020,hill_2021}---we ask whether contemporary superconducting processors can be tailored for performing efficient stabilizer quantum error correction. Such a strategy would reduce the requirements for fault-tolerance while taking advantage of the significant investment towards machines for near-term quantum applications. While other recent advances in optimizing the surface code have focused on using qubit pair measurements to eliminate the use of two-qubit entangling gates~\citep{Gidney_2022,Chao_2020}, we take a different approach by implementing a single five-qubit entangling gate. Notably, ion traps have successfully utilized a native multi-qubit gate, the M{\o}lmer-S{\o}rensen (MS) gate~\citep{Molmer1999}, for efficient stabilizer measurements in error correction experiments~\citep{Erhard_2021,Nigg_2014}. Moreover, there is numerical evidence that error correction thresholds do not suffer from the more complex noise associated with the multi-qubit MS gate~\citep{Bermudez_2017,Schwerdt_2022}.

In the following, we show how to adapt the control of contemporary gate-model superconducting processors to achieve a different error profile, relative to standard quantum circuit constructions. Our superconducting processor layout is based on planar transmons arranged in a square array with one qubit coupled to four nearest-neighbors via tunable couplers, which is a leading design for near-term processors~\citep{Google_supremacy,Foxen_2020,Will_Oliver_coupler_2021,Sete2021}. To operate the chip for error correction (see Fig.~\ref{fig:hop_overview}), exchange coupling rates and qubit frequencies are tuned to engineer a strong-dispersive $ZZ$ interaction pairwise between data qubits and local stabilizer check qubits constant during the gate\footnote{This interaction is the dual of the MS gate for ions, which is generated by an $XX$ interaction~\citep{Molmer1999}. We therefore anticipate that related analysis of MS-based codes~\citep{Bermudez_2017,Schwerdt_2022} will carry over to a large degree.}. Evolved over time, the interaction generates a Hardware Optimized Parity (HOP) gate, controlled on the data register and targeting a local stabilizer check qubit---mapping the odd and even parity states of the register onto two states of the stabilizer check qubit. Measuring the stabilizer qubit then records the associated error syndrome. The HOP-based design reduces the system calibration problem to a single gate per check.

Our analysis considers HOP gates performing the surface codes with a realistic noise model that captures incoherent errors, measurement errors, as well as spurious next-nearest neighbor Hamiltonian terms. Numerical simulations indicate that the HOP surface code does not correct the same number of errors as a traditional surface code with optimized $\textrm{CNOT}$ schedule. Surprisingly, however, the HOP design may have error correction advantages for near-term machines due to a higher threshold. As a result, we estimate modest physical overhead reductions ($15\%$) are possible, for example, at the scale needed for a target application in pricing financial derivatives~\citep{Chakrabarti_2021}.

This manuscript is organized into three technical sections. First, we show how a superconducting processor being developed for gate-model applications can be tuned to support efficient HOP gates. Second, we discuss the associated device noise model for a HOP gate. Finally, we analyze the performance of rotated surface codes mediated by HOP gates, relative to standard constructions.
\section{Superconducting Processor Design}
\label{sec:dev_design}
In the following section, we discuss the physical mechanisms underpinning the HOP gate using superconducting qubits. This includes both a micro-architecture based on transmon circuits with tunable couplers as well as system design considerations towards heat dissipation. Overall, we seek to realize a five-qubit HOP gate via a pairwise dispersive interaction between a single stabilizer qubit and four local data qubits with equal coupling strength,
\begin{equation}\label{eq:H_disp}
    H_{\text{disp}} = \frac{-\zeta(t)}{4} \sum^{4}_{k=1} Z_0 Z_k,
\end{equation}
where $\zeta(t)$ is a time-dependent interaction strength (described in more detail soon), $Z_k$ is the Pauli-Z operator acting on the $k$-th qubit, with the index $k=0$ assigned to the stabilizer check qubit. Allowing the five-qubit system evolve under this Hamiltonian with a constant interaction strength $\zeta_0$ for a time $\tau=\pi/\zeta_0$ corresponds to applying a total unitary $U=\exp{i \pi/4 \sum_k Z_0 Z_k}$ which is equivalent to the four CZ gate construction shown in Fig.~\ref{fig:hop_overview} up to single-qubit phase gates applied to the data qubits ($\sqrt{Z}$ rotations on each).  Within a protocol for $Z$-stabilizer measurements (see Fig.\ref{fig:hop_overview}), the system is biased with no interactions ($\zeta(0)=0$), and in this configuration, stabilizer qubits are prepared along the $X$-axis via Hadamard ($H$) gates. After turning on the interaction and waiting an amount $\tau$, the interaction is turned off $\zeta(t\geq\tau)=0$. A final Hadamard gate maps the state of the stabilizer qubit as $\langle Z_0 \rangle = -1$ for even number of excited data qubits, and $\langle Z_0 \rangle = +1$ for odd. Thus, HOP gates enable a single-operation parity check on the four data qubits. For the typical gate-model configurations we consider here, estimates of $\zeta/2 \pi= 5\ \text{MHz}$ corresponds to a HOP gate time of $\tau=100~\textrm
{ns}$. Measuring $X$-stabilizers instead has $H$ gates applied to the data qubits but otherwise follows the same protocol.

\begin{figure}
    \centering
    \includegraphics[width=6cm]{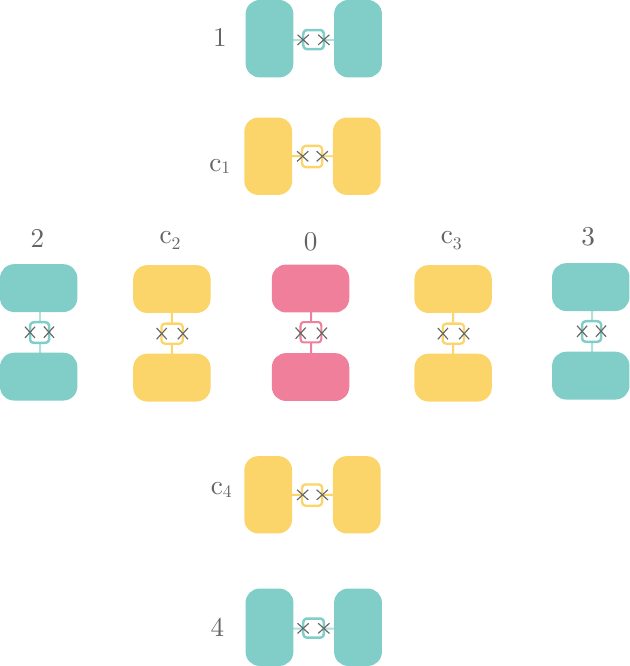}
    \caption{Schematic of a 5-qubit lattice for four-qubit parity check measurement. Stabilizer qubit (magenta) is coupled to four data qubits (teal) via asymmetric floating tunable couplers  $c_j$ (yellow) introduced in Ref. \cite{Sete2021}.}
    \label{fig:connectivity}
\end{figure}

The Hamiltonian \eqref{eq:H_disp} can be realized by coupling the stabilizer qubit to the data qubits using a tunable coupler, following the schematic approach shown in Fig.~\ref{fig:connectivity}. In this design, we utilize the asymmetric tunable coupler introduced in \cite{Sete2021}. For this type of coupler the zero-coupling condition is achieved by placing the tunable coupler frequency above that of the qubits. The qubit-qubit coupling can be varied from $\zeta=0$ to the target coupling $\zeta_0$ by tuning the flux bias of the tunable coupler. In gate-model operation, the couplings between the qubits are turned on and off to actuate two-qubit entangling gates. However, here we activate multiple couplings simultaneously.

Following \cite{Sete2021}, the five qubit and four tunable coupler system can be described by the following Hamiltonian with a three-level approximation for each anharmonic mode
\begin{align}\label{eq:q5}
    &H = \sum_{j=0}^4\omega_j(\Phi_j)|1\rangle_j\langle 1| + (2\omega_j(\Phi_j)-\eta_j(\Phi_j))|2\rangle_j\langle 2|\notag\\
    &+\sum_{k=1}^{4}\omega_{c_k}(\Phi_k)|1\rangle_{c_k}\langle 1| + [2\omega_{c_k}(\Phi_k)-\eta_{c_k}(\Phi_k)]|2\rangle_{c_k}\langle 2|\notag\\
    & + \sum_{m=1}^{4}\left[g_{0m}X_0X_m + g_{0c_m}X_0X_{c_m}+ g_{mc_m}X_mX_{c_m}\right],
\end{align}
where $\omega_j$ and $\eta_j$ are the $j$th qubit frequency and anharmonicity; $\omega_{ck}$ and $\eta_{ck}$ are frequency and anharmonicity of the $k$th tunable coupler, $g_{0m}, (m\in [1,4]$) are the direct stabilizer qubit-data qubit couplings, $g_{0c_m}$($m\in [1,4]$) are the stabilizer qubit-tunable coupler couplings, and $g_{mc_m}$($m\in [1,4]$) are the data qubit-tunable coupler couplings. Here $X_{j} = \sigma_{-,j} +\sigma_{+,j}$, where $\sigma_{-,j} = |0\rangle_j\langle 1| +\sqrt{2} |1\rangle_j\langle 2|$ is the lowering operator for the $j$th qubit or coupler. We assume the qubits and tunable couplers are flux tunable transmon qubits and their frequencies as a function of the flux bias can be approximated by
\begin{align}
    \omega_{m}(\Phi_m)& = \sqrt{8 E_{J\mathrm {eff},m}E_{C,m}}-E_{C,m}(1+\xi_{m}/4 +21\xi_{m}^2/128),\\
    \eta_{m}(\Phi_m) &= E_{C,m}(1 + 9 \xi_{m}/16 + 81 \xi_{m}^2/128),\\
    E_{J\mathrm {eff},m} &= \frac{E_{J,m}}{1+r_m} \sqrt{1+r_m^2+2r_m\cos(2\pi \Phi_m/\Phi_0)},
\end{align}
where $m\in \{[1,4],[c_1,c_4]\}$, $\xi_m = \sqrt{2E_{C,m}/E_{J\mathrm {eff},m}}$ and $r_m = E_{J1,m}/E_{J2,m}$ is the ratio of the junction energies, $E_{J1,m}, E_{J2,m}$ of the SQUID, $E_{C,m}$ is the charging energy, and $E_{J,m} = E_{J1,m}+E_{J2,m}$ is the total junction energy; $\Phi_m$ and $\Phi_0$ are the external flux bias and flux quantum, respectively. Note that the qubit-coupler couplings $g_{0c_m}$ and $g_{mc_m}$ depend on the external flux biases $\Phi_m$. Explicit  dependence of the these couplings are used when computing the nearest neighbor $(Z_0 Z_k)$ and next-nearest neighbor $(Z_{j\neq 0} Z_k)$ couplings. The designed operating point assumes flux-biasing to adjust all qubits to their maximum frequencies and all tunable couplers to their minimum frequencies ($\Phi_{k} =0.5\Phi_0$) in order to maximize the dispersive stabilizer-data qubit couplings while keeping the $Z_{j\neq 0} Z_k$ coupling minimum, although small bias adjustments are anticipated to account for fabrication tolerances. 

We estimate the coupling strengths numerically by exact diagonalization of the Hamiltonian comprising three energy levels for each of the five transmons and four tunable couplers ($3^9$ dimensions), on a workstation with 256~GB of RAM and 128 CPU cores. We identify the dispersive coupling strengths for each interaction as the frequency difference, 
\begin{equation}
    \zeta_{jk} = \omega_{|11\rangle_{jk}}+ \omega_{|00\rangle_{jk}} -\omega_{|10\rangle_{jk}}-\omega_{|01\rangle_{jk}}
\end{equation}
where $\omega_{|nm\rangle_{jk}}$ is the eigenfrequency associated with $n$ ($m$) excitations in the $j$-th ($k$-th) mode and $-\zeta_{jk}/4$ is the weight of the $Z_j Z_k$ term in the interaction Hamiltonian as in Eq.~\ref{eq:H_disp}. For realistic parameters for asymmetric tunable couplers and qubits (shown in Table~\ref{tab:couplings} and Table~\ref{tab:freqs}), we find that $\zeta_{0k}/2\pi=5.0$ MHz and $\zeta_{j k}/2\pi= 2-24$ kHz for $j\neq 0$. These parameters are used for the construction of the following error model and subsequent fault-tolerant analysis.

\section{Error Model}
\label{sec:error_model}
We set out to provide a realistic description of the noise model associated with the HOP gate that is compatible with fault-tolerant analysis. A noisy quantum process can be described as a superoperator $\mathcal{E}(\rho)$ that takes an initial density matrix $\rho$ to a final density matrix~\citep{nielsen00}. Here, we are interested in describing the error channels associated with the HOP gate, rather than the more cumbersome description of the full process itself. We therefore write our circuit as the composition of the ideal unitary channel $U$ and an error channel $\Lambda$:
\[
\Qcircuit @C=.5em @R=0em @!R {
& \gate{\mathcal{E(\rho)}} & \qw & \push{\rule{.3em}{0em}=\rule{.3em}{0em}} & \qw & \gate{U} & \gate{\Lambda} & \qw 
}
\]
\noindent with time proceeding from left to right. Formally, that basis change is conducted via:
\begin{equation}
    \mathcal{E}(\rho) = \Lambda \circ U(\rho) = \sum_{i,j=0}^{d^2-1} \chi^{\Lambda}_{ij}\mathcal{P}_i U \rho U^{\dagger}\mathcal{P}_j^{\dagger},
\end{equation}
\label{eq:twirl}
\noindent where $\chi^{\Lambda}$ is an error process matrix~\citep{Korotkov2013}; and $\{\mathcal{P}_k\}$ are elements of the multi-qubit Pauli operators, $\mathcal{P}_k\in\{I,X,Y,Z\}^{\otimes n}$. This allows us to look at errors in an interaction picture. For instance, the ideal unitary channel: $\mathcal{E}(\rho)=U\rho U^{\dagger}$ has only one non-zero element $\chi^{\Lambda}_{00}=1$ with $\mathcal{P}_{00}=I^{\otimes n}$. 

To study the effect of noise in the context of HOP surface codes, where stabilizer-type simulations are desired, we include noisy single-qubit Pauli gates\footnote{Added noise is discussed Sec.~\ref{sec:ec_sims}.} to perform Pauli-twirling of $\Lambda$ \citep{bennett_mixed-state_1996, bennett_purification_1996, geller_efficient_2013}, with the twirled error channel denoted as $\tilde{\Lambda}$. In the twirling approximation, the twirled error process matrix $\chi^{\tilde{\Lambda}}$ is diagonal with eigenvalues $p_k$ each of which is a real-valued number. The weight of $p_k$ is the probability that the $\tilde{\Lambda}$ channel applies the $\mathcal{P}_k$ rotation, instead of the desired $I^{\otimes n}$ element. Following the convention in Eq.~\ref{eq:twirl}, these probabilistic Pauli gates occur after the perfect unitary. 

To estimate $\chi^{\Lambda}$, we solve the Lindblad master equation (ME)~\citep{manzano_short_2020, nielsen00} governing the HOP gate under the types of Markovian decoherence that are the main noise in superconducting qubits, namely energy loss $T_1$ processes and pure dephasing $T_{\phi}$ processes~\citep{Krantz_2019}. These act concurrently with interaction Hamiltonian of the gate Eq.~\ref{eq:H_disp} to evolve the density matrix as
\begin{equation}\label{eq:ME}
	\begin{split}
    	\frac{d \rho}{d t} &= -i [H,\rho] + \sum_j \left[L_j \rho L^{\dagger}_j - \frac{1}{2} L_j  L^{\dagger}_j \rho - \frac{1}{2} \rho L_j  L^{\dagger}_j \right] \\
		&\equiv \mathcal{L}(t)\rho(t)
	\end{split}
\end{equation}
where $L_j$ are Lindblad jump operators, and $\mathcal{L}$ is an infinitesimal generator of the evolution sometimes called the Liouvillian superoperator \citep{manzano_short_2020, nielsen00}. 

While we will soon allow for decoherence acting each of the five qubits in the HOP gate, for the time being, we consider the effect of noise on the parity check only, which can help develop an intuition for the subsequent full analysis. In this case, the Lindblad jump operators are
\begin{equation}\label{eq:single_lindblad_op}
	L_j \in \left\{\sqrt{\Gamma_1}\sigma_{-}\otimes I^{\otimes 4},\sqrt{\Gamma_{\phi}}Z\otimes I^{\otimes 4}\right\},	
\end{equation}
where $\Gamma_1 = 1/T_1$ and $\Gamma_{\phi} = 1/T_{\phi}$ for the parity check qubit, and $I^{\otimes 4}$ acts on the data qubit register. This single-qubit noise generates multi-qubit back-action on the data qubits. To see this, it is helpful to consider Eq.~\ref{eq:ME} in the rotating frame of the interaction Hamiltonian (Eq.~\ref{eq:H_disp}), as
\begin{equation}
    \tilde{\rho}(t) \equiv e^{iHt}\rho(t)e^{-iHt}
\end{equation}
and the jump operators similarly get transformed
\begin{align}
    \tilde{\sigma}_{-}(t)&=e^{iHt}\sigma_{-}e^{-iHt}=\exp{-i\frac{\zeta t}{4} \sum_{k} Z_k}\sigma_{-}  \\
   \nonumber \tilde{Z}(t) &= e^{iHt}Z e^{-iHt} = Z.
\end{align}
Energy decay of the parity check qubit generates a phase kick onto the data qubits that is time-dependent, amplifying the effect of that error in the HOP gate. On the other hand, a dephasing event commutes through the Hamiltonian during the actuation of the HOP gate, and therefore, remains a single-qubit error. Similarly, when we include single-qubit decoherence on the data qubits, energy relaxation from these qubits gives a phase kick to the parity check qubit.

\begin{figure}
	\includegraphics[width=\columnwidth]{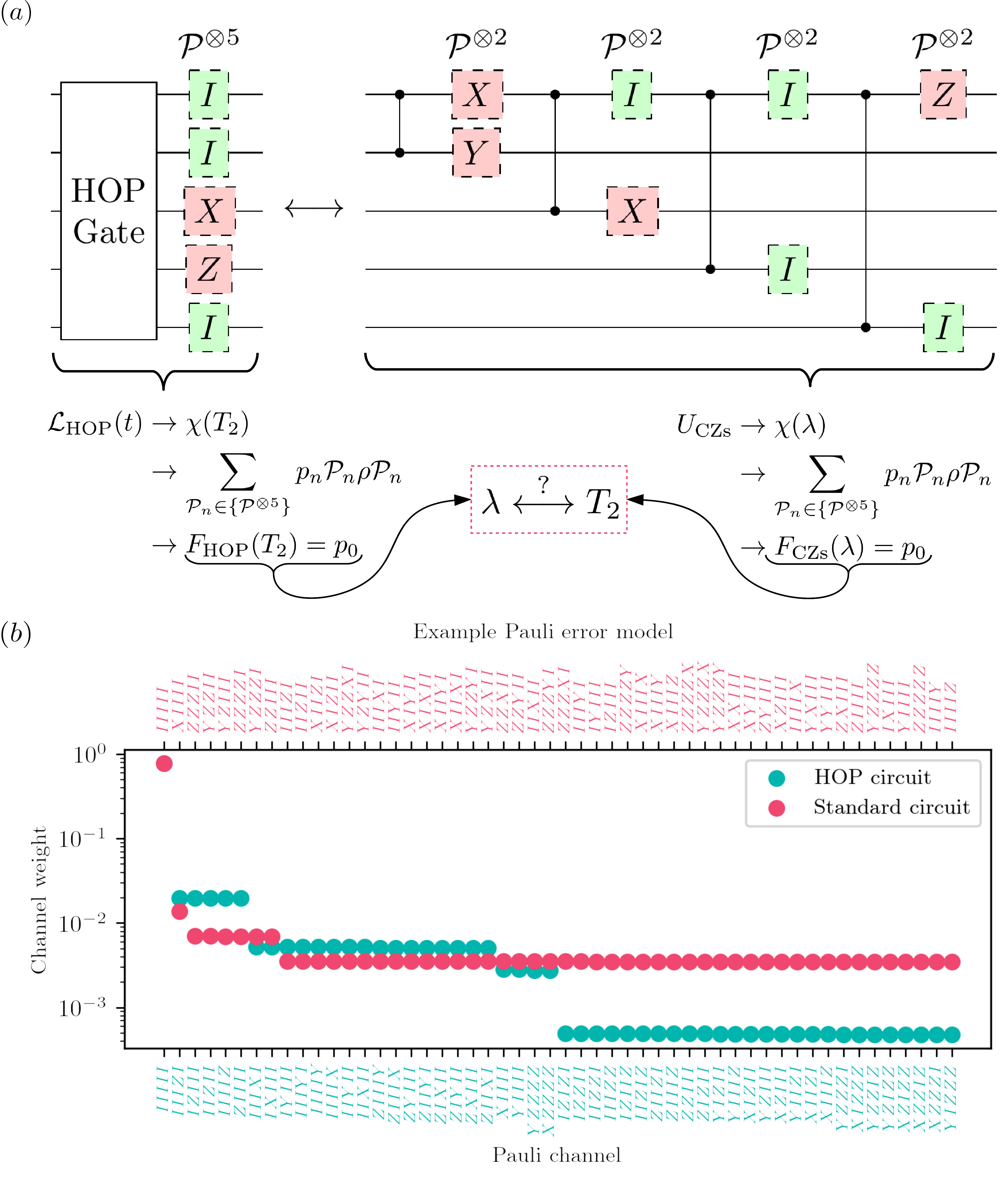}
	\caption{\label{fig:error_model} (a) Comparing the five-qubit Pauli error model for the HOP gate to a corresponding two-qubit depolarizing error model for the standard circuit, illustrated by an unwanted $\mathcal{P}\in\mathcal{P}^{\otimes 5}$ error for the HOP gate and unwanted $\mathcal{P}\in\mathcal{P}^{\otimes 2}$ errors after every two-qubit gate for the standard circuit. The Louiviallian superoperator for a full HOP gate $\mathcal{L}_{\text{HOP}}(t)$ is used to get the error process matrix $\chi$ parameterized by coherence times $T_2$, from which we get an approximated error model made up of five-qubit Paulis $\mathcal{P}=\{I,X,Y,Z\}^{\otimes 5}$. The equivalent operation made up of four CZ gates each undergoing two-qubit depolarizing noise is used to produce a superoperator process unitary $U_{\text{CZs}}$, from which we get the process matrix for an effective five-qubit Pauli channel parameterized by a two-qubit depolarizing rate $\lambda$. We equate the process fidelities $F_{\text{HOP}}(T_2) = F_{\text{CZs}}(\lambda)$ to determine an equivalence between $\lambda$ and $T_2$. (b) Resulting error models from the analysis in (a) using $\Gamma_2 = 1/T_2 = 0.05$ and a corresponding $\lambda \approx 0.066$, for channels whose weight has probability greater than $3\times 10^{-4}$ (arbitrary cutoff chosen for fitting to plot), comparing the effective error channels for both equivalent circuits. Note the process fidelities given by $IIIII$ are equal.}
\end{figure}
We turn to numerics to solve the ME with equal decoherence rates across all qubits. Specifically, we use QuTiP~\citep{johansson_qutip_2013} to find the propagator $V_t$, which is a superoperator describing the full noisy channel, as $V_t \left(\rho(0)\right)=\rho(t)$, by numerically integrating the Liouvillian:
\begin{equation}\label{eq:v_t}
    V_t = \mathcal{T} \exp{\int_0^t ds \mathcal{L}(s)},
\end{equation}
where $\mathcal{T}$ is the time-ordering operator. The time $\tau$ for the evolution is chosen such that the interaction Hamiltonian achieves the correct entangling phase ($\tau=\pi/\zeta_0$). To estimate the error process matrix $\chi^{\Lambda}$, we perform tomography on the composite channel: $V_t \circ U^{\dagger}$ with $U$ the target multi-qubit phase gate, $U=\exp{i(\pi/4)\sum_k Z_0 Z_k}$. Pauli twirling removes the off-diagonal terms of $\chi^{\Lambda}$. That yields $\chi^{\tilde{\Lambda}}$, from which we obtain the weights $p_k$ of the corresponding Pauli channels $\mathcal{P}_k$. Succinctly, this procedure allows us use a low-level description of our system from Eq.~\ref{eq:H_disp} and Eq.\ref{eq:v_t} to establish description of the error model of the form of
\begin{equation}
    \tilde{\Lambda}_{\text{HOP}}(\rho) = \sum_{i=0}^{1023}w_i\mathcal{P}_i\rho\mathcal{P}_i,\ \mathcal{P}_i \in \{I, X, Y, Z\}^{\otimes 5},
\end{equation}
\label{eq:kraus_model}
\noindent which we analyze with a parameterized noise model: we take a physical error rate $p$ for our fault-tolerance calculations given by the HOP gate time and the single-qubit decoherence rate as $p = \tau \Gamma_2$, with $\Gamma_2 = \Gamma_1/2 + \Gamma_{\phi}$. To reduce the simulation overhead, we make the further assumption that $\Gamma_{\phi}=\Gamma_1/2$ ($T_1 = T_2$), which is typical of superconducting qubits~\citep{Krantz_2019}, and ignore leakage states. With this approach, we can produce a more complete error model, as illustrated in Fig.~\ref{fig:error_model}(b) that includes energy loss and dephasing on all five qubits\footnote{Additionally, we have used this framework to investigate undesired effects including the coherent $ZZ$ cross-talk anticipated for next-nearest neighbors in Sec.~\ref{sec:dev_design} by adding these terms to the ME. Exploring the effects using parameters guided by simulated results (e.g. App.~\ref{app:nnn}), we found no discernible impact on the resulting error model and subsequent logical error threshold calculations. Thus, for simplicity, we report in the following section only results from the decoherence model.}.

Ultimately, we are interested in understanding how the HOP gate performs in the context of error correction. Towards that end, we develop a comparison point based on a simplified quantum channel construction using the logically equivalent four CZ gates (see Fig.~\ref{fig:hop_overview}), which we call the standard construction. For the standard construction, we assume two-qubit deploarizing noise applied after each CZ gate:
\begin{equation}
    \Delta_{\lambda}(\rho) = \sum_{i=0}^{15}v_iK_i\rho K_i
\end{equation}
where the operators $K_i$ are two-qubit Pauli operators $\mathcal{K}_i\in\mathcal{P}^{\otimes 2}$ with the corresponding weights
\begin{equation}
    v_0 = 1-\frac{15\lambda}{16},\ v_{k\neq0} = \frac{\lambda}{16}.
\end{equation}
Some of these depolarizing errors are amplified in the four CZ circuit, resulting in a non-trivial five-qubit error process matrix $\tilde{\Lambda}_{\text{CZ}}$ with an equivalent form as Eq.~\ref{eq:kraus_model}. 

Recognizing that the overall process fidelity is $\tilde{\Lambda}^{00}_{\text{HOP}} = F_{\text{HOP}}(T_2)$ and $\tilde{\Lambda}^{00}_{\text{CZ}} = F_{\text{CZ}}(\lambda)$, we set $F_{\text{HOP}}(p) = F_{\text{CZ}}(\lambda)$ to determine what two-qubit depolarizing rate $\lambda$ is comparable to $p$. Using this procedure for determining equal fidelity (used in Sec.~\ref{sec:ec_sims} set a consistent Physical Error Rate) produces the error models in Figure~\ref{fig:error_model}(b), where we note that the channel weights for the Pauli $I^{\otimes 5}$ are the same. The differences in the distribution of errors across the Pauli strings illustrates the non-triviality of the relationship between these two logically equivalent processes. We observe that the most prominent Pauli channels are different between the two methods. Moreover, nearly half of Pauli channels for the HOP gate are an order of magnitude smaller than for the standard circuit, meaning the HOP based gate error may be simpler to characterize and operate in practice, a surprising potential advantage of the multi-qubit gate.

\section{Error Correction Simulations}
\label{sec:ec_sims}
In this section, we examine fault-tolerant error correction via the rotated surface code using optimized syndrome measurement circuits based on HOP gates. In particular, we compare logical failure rates and the fault-tolerant threshold achievable using HOP syndrome measurement circuits against standard syndrome measurement circuits based on two-qubit entangling gates. We also compare resources needed for either scheme to implement a logical qubit with a target logical failure rate of $p_L=10^{-10}$ \cite{Chakrabarti_2021}. Logical failure rates are estimated via Monte Carlo simulations for a single fault-tolerant quantum error correction cycle. We empirically observe a fault-tolerant threshold for the HOP surface code that is approximately 1.5 times higher than the threshold estimated for the standard surface code. This suggests that the multi-qubit correlated error mechanisms present in the 5Q and 3Q gates are not strong enough to negatively affect the efficacy of this scheme at near-threshold error rates. We compare resource requirements for using either scheme to implement a logical qubit with logical error rate $10^{-10}$ and find that in regimes of interest the HOP scheme does not require greater spacetime resources than the standard scheme and may be preferable based on hardware considerations.

\subsection{Parity Check Circuits and Error Models}

In a typical surface code architecture, a single fault-tolerant error correction cycle is made up of $d$ consecutive rounds of syndrome measurements, where $d$ is the distance of the surface code patch encoding the logical qubit \cite{Fowler_2012}. Each round of syndrome measurements consists of a single measurement of each $X$ and $Z$-type stabilizer generator of the code. Repeated rounds of syndrome measurements aid in offsetting the effects of measurement errors which corrupt the measured syndrome.

\begin{figure}
\centering
\includegraphics[width=0.5\textwidth]{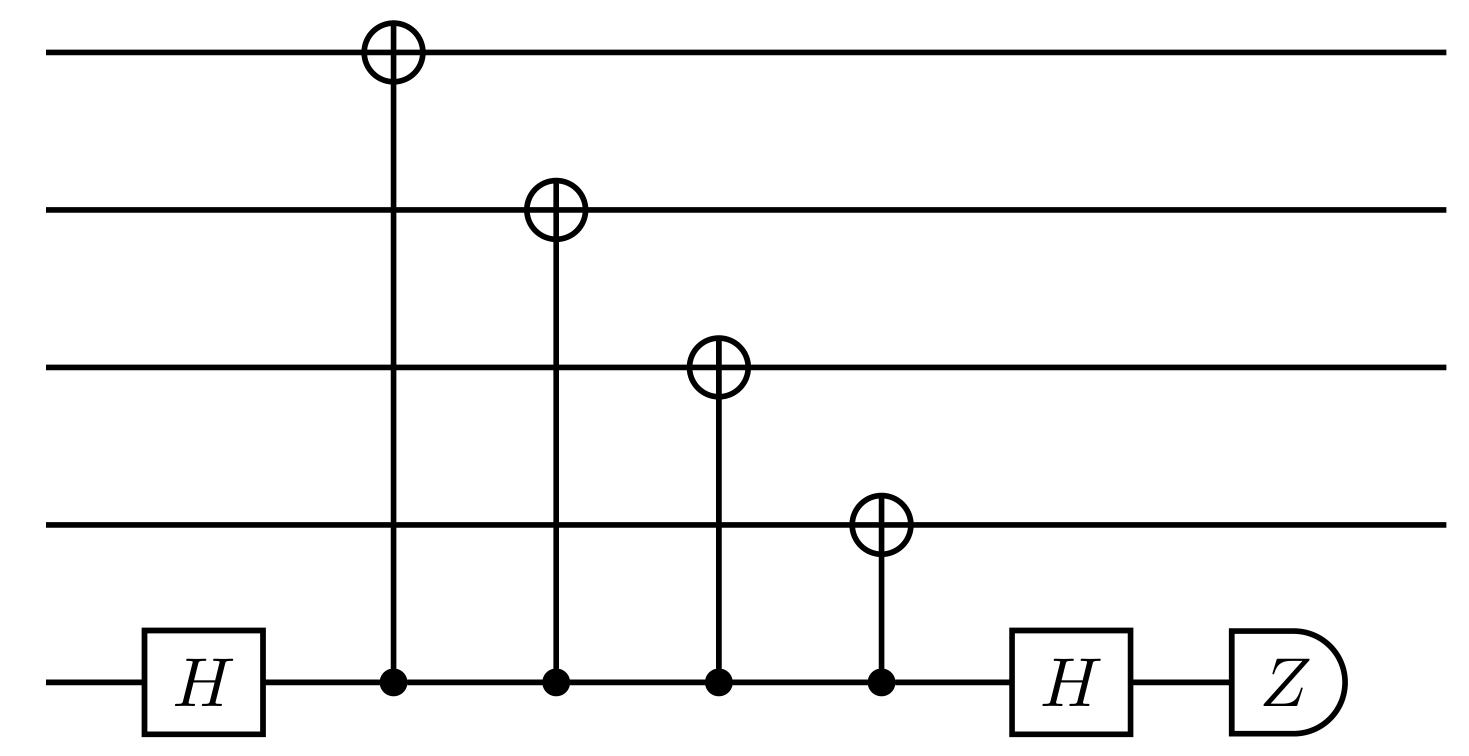}
\caption{\label{fig:checkcirc}Standard implementation of an $X$-type parity check in the surface code. The first four qubits are data qubits of the code state participating in the check, and the fifth qubit is an ancilla qubit whose measurement produces the syndrome bit for this parity check. A $Z$-type check can be executed by replacing the $\textrm{CX}$ gates with $\textrm{CZ}$ gates. Note that this circuit has a time depth of 6.}
\end{figure}

\begin{figure}
\centering
\includegraphics[width=0.8\textwidth]{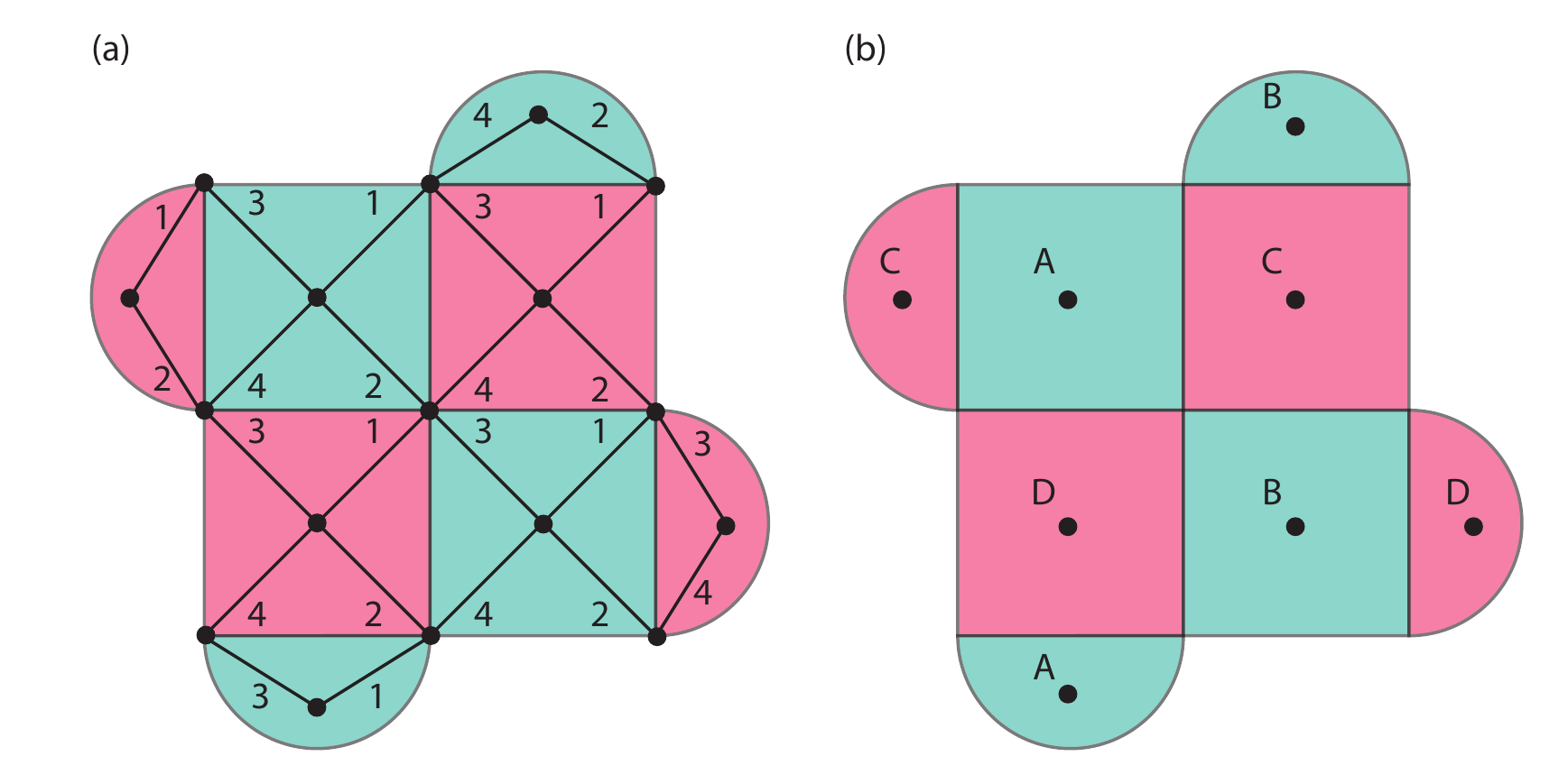}
\caption{\label{fig:abcd}Schedules for surface code operation. (a) Optimal scheduling of two-qubit gates that allow simultaneous measurement of all parity checks of a distance 3 rotated surface code. Teal squares represent $Z$-checks, magenta squares represent $X$-checks, black circles on vertices represent data qubits, black circles on faces of tiles represent check qubits, and diagonal lines denote a 2-qubit gate ($\textrm{CX}$ or $\textrm{CZ}$) coupling a data and check qubit. Performing the 2-qubit gates in the order indicated allows all stabilizer measurements to be carried out simultaneously while avoiding certain fault mechanisms that could reduce the effective distance of the code. (b) Scheduling of groups of parity checks carried out using multi-qubit parity check HOP gates. Groups of parity checks are carried out in the order ABCD.}
\end{figure}

Individual stabilizer generators are measured non-destructively by coupling the relevant data qubits to a single ancilla qubit which is then measured to infer the outcome. Typically this is achieved using CX or CZ gates between the data qubits and the measurement ancilla \cite{Fowler_2012} (see Figure \ref{fig:checkcirc}). The surface code architecture allows for all stabilizer measurements in a single round to be carried out simultaneously in parallel when using two-qubit entangling gates (see Figure \ref{fig:abcd}a). By mandating a specific schedule for the order in which the two-qubit gates are executed, one can avoid reducing the effective distance of the code by spreading $X$ or $Z$ errors along their respective logical operator directions. We refer to this circuit described using Figures \ref{fig:checkcirc} and \ref{fig:abcd}a as the standard parity check circuit for the surface code.

\begin{figure}
\centering
\includegraphics[width=0.8\textwidth]{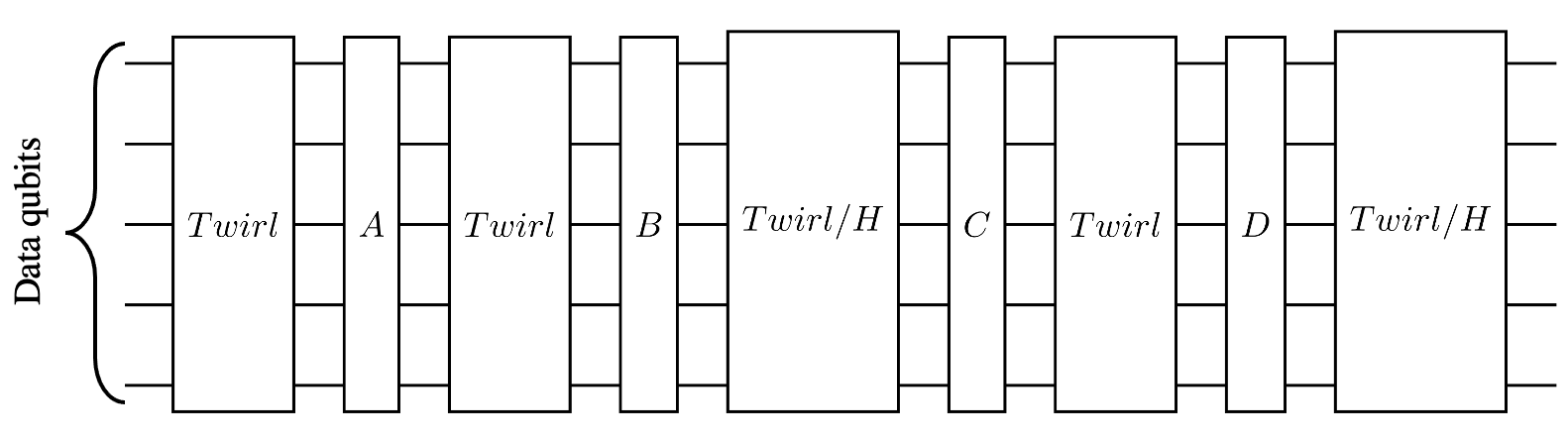}
\caption{\label{fig:rough}Circuit depicting a single round of stabilizer measurements using multi-qubit parity check gates. Blocks labeled A/B/C/D include the preparation of ancillas, coupling of the ancillas and data qubits via the appropriate multi-qubit parity check gates, and measurement of the ancillas. Blocks labeled 'Twirl' consist of random single qubit Paulis applied to the data qubits in order to justify the twirling approximation discussed in Sec.~\ref{sec:error_model}. Blocks labeled 'Twirl/H' also include Hadamard gates on the data qubits required in order to execute $X$-checks, with the assumption that the Hadamard gates can be compiled into the Twirling steps when required. }
\end{figure}

Now we describe an alternative syndrome measurement circuit for the HOP surface code. We refer to this as the HOP circuit. One apparent benefit of using the multi-qubit interaction gates to perform parity checks instead of two-qubit gates is that a single parity check can effectively be performed faster. However, when using the multi-qubit interaction gates, one loses the ability to perform all parity checks in parallel as was possible with the standard circuit. Extra time before and after each multi-qubit gate is also required in order to include the twirling steps that justify our error model (Sec.~\ref{sec:error_model}). In Figures \ref{fig:abcd} and \ref{fig:rough}, we present a schedule for carrying out an entire round of parity checks using the HOP gates across 9 time steps.

The circuit-level error models that we use to simulate error correction using either circuit are characterized by the single parameter $p$ described in Sec.~\ref{sec:error_model}. Any single qubit gate location (Hadamard, Pauli twirl or idle/identity) is followed by a single qubit depolarizing channel of strength $p_1=p/10$. Ancilla preparation and measurement locations fail with probability $p_{pm}=p/2$, where a preparation failure causes the preparation of the state orthogonal to the one intended, and a measurement failure flips the measurement outcome. For the standard circuit, the failure rate of the two-qubit gates, $p_2$, is set by demanding that the process fidelity of performing a full single parity check using two-qubit gates is equal to the process fidelity of performing the parity check with the 5-qubit HOP gate. This allows us to characterize the effects of high-weight correlated errors from the HOP gates relative to the lower weight errors caused by failures of two-qubit gates. All A/B/C/D blocks in Figure \ref{fig:rough} are followed by the correlated multi-qubit noise channels (Sec.~\ref{sec:error_model}) on the qubits that participated in those checks.

\subsection{Simulations}

We use Stim \citep{Gidney_2021} in order to generate Monte Carlo samples of fault patterns and their associated syndrome patterns in our syndrome measurement circuits assuming $d$ noisy rounds of syndrome measurement followed by one perfect round to project back to the codespace \citep{Dennis_2002}. For each sample, we feed the results to a minimum-weight perfect matching (MWPM) decoder implemented in Python using PyMatching \citep{PyMatching} and record whether or not the decoder's correction causes a logical failure by flipping a hypothetical measurement outcome of the logical Z or X operators of the code. Because Stim does not support the use of correlated Pauli channels acting on more than two qubits, the multi-qubit correlated channels are approximated in our simulations by using appropriately chosen single outcome Pauli channels as described in the appendix. Under this approximation, the probabilities of any correlated Pauli error occurring after a multi-qubit interaction gate differ from the probabilities intended by the true error model deviate only by corrections that are $O(p^3)$, which should be irrelevant in the low $p$ regime.

\begin{figure}
\centering
\includegraphics[width=0.8\textwidth]{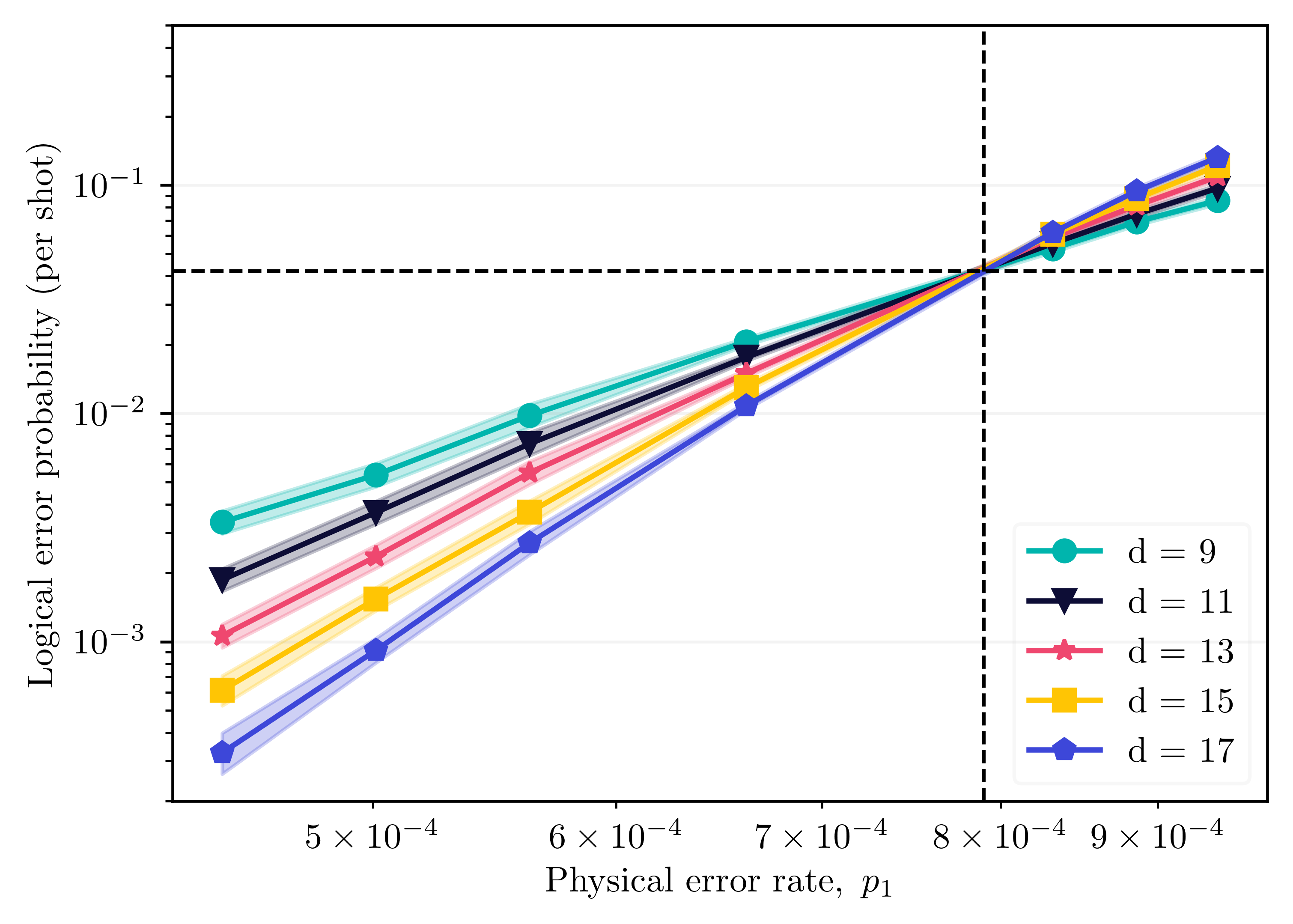}
\caption{\label{fig:StdThresh}Logical failure rate vs physical single qubit failure rate $p_1$ for various distances $d$ of the standard rotated surface code syndrome measurement circuit. Each data point reflects either $10^6$ Monte Carlo trials, or the number of trials required to observe $10^4$ logical failures, whichever is less. We observe that the threshold for the standard circuit under the described error model is approximately $0.79\times 10^{-3}$. Shaded areas around curve represent a 99.9\% confidence interval computed via binomial relative likelihood ratios.}
\end{figure}

\begin{figure}
\centering
\includegraphics[width=0.8\textwidth]{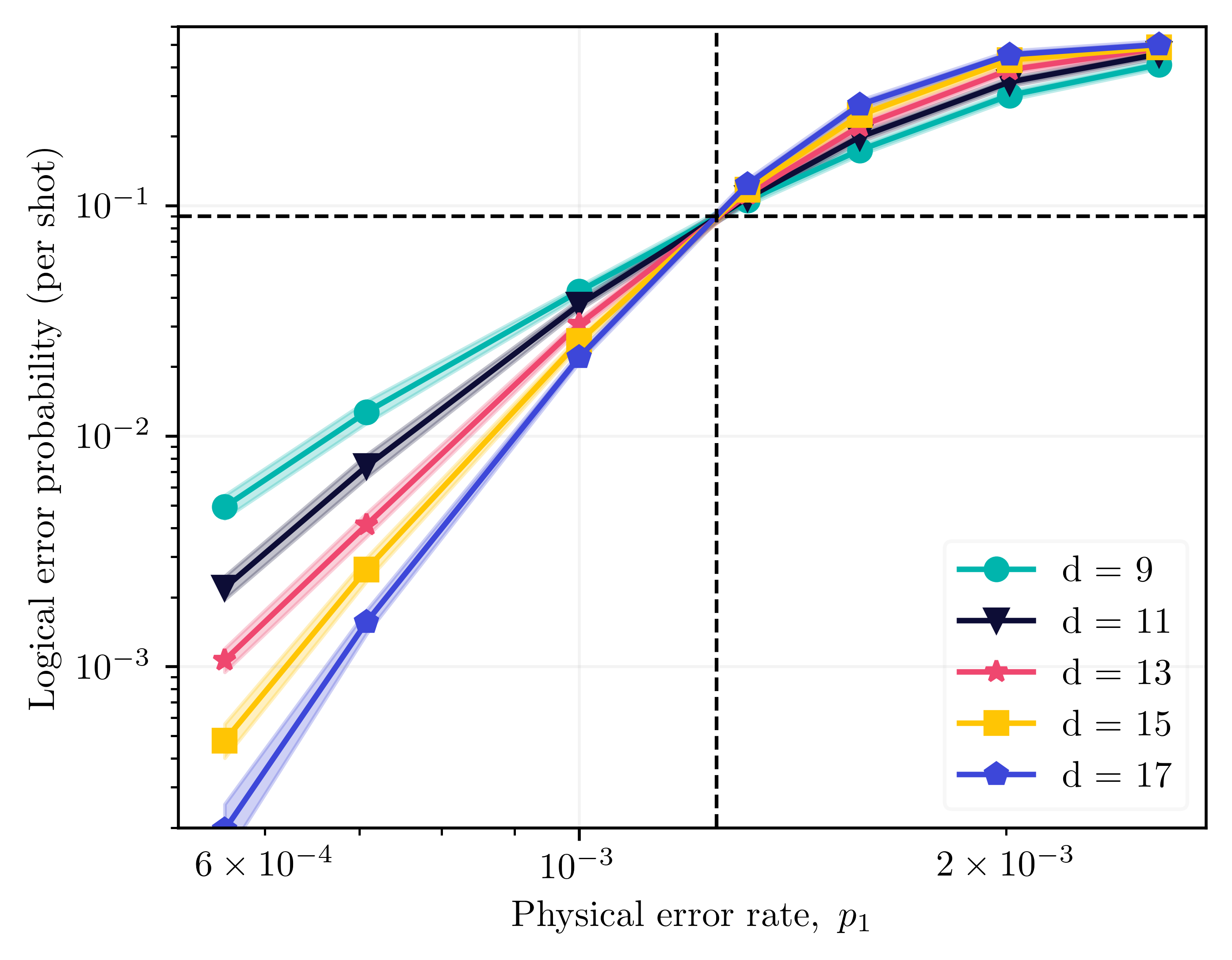}
\caption{\label{fig:5QThresh}Logical failure rate vs physical single qubit failure rate $p_1$ for various distances $d$ of the HOP syndrome measurement circuit. Each data point reflects either $10^6$ Monte Carlo trials, or the number of trials required to observe $10^4$ logical failures, whichever is less. We observe that the threshold for the standard circuit under the described error model is approximately $1.25\times 10^{-3}$. Shaded areas around curve represent a 99.9\% confidence interval computed via binomial relative likelihood ratios.}
\end{figure}

In Figures \ref{fig:StdThresh} and \ref{fig:5QThresh}, below, we see simulation results for the standard and HOP circuits respectively. We observe a threshold of approximately $1.25 \times 10^{-3}$ for the HOP circuit and a threshold of approximately $0.79 \times 10^{-3}$ for the standard circuit. As such, the HOP scheme appears to have an empirical threshold 1.5 times higher than that of the standard scheme. Note that our observed threshold for the standard scheme appears to be roughly an order of magnitude lower than the typically reported range of roughly 0.005 to 0.01 for the surface code under circuit-level noise \citep{Stephens_2014}. We attribute this discrepancy to our use of different failure rates for single qubit gates, multi-qubit gates, preparations, and measurements.

\subsection{Resource Estimates}

Here we compare the physical resources required to use either error correction scheme to implement a logical qubit encoded in a rotated surface code patch with a target logical failure rate of $p_L=10^{-10}$. We choose this logical error rate motivated by a recent study indicating that roughly $10^{10}$ logical operations would be required to execute a quantum algorithm for derivative pricing in a regime where quantum advantage could be achieved \cite{Chakrabarti_2021}. In order to compare resources, we fit curves describing logical failure rate as a function of physical error rate $p_1$ below threshold for various distances, and extrapolate those curves down to find the physical failure rate needed in order to achieve $p_L=10^{-10}$ for each distance requiring a different number of physical qubits to implement. Linear fit data used to perform these extrapolations can be found in the appendix.

\begin{figure}
\centering
\includegraphics[width=0.8\textwidth]{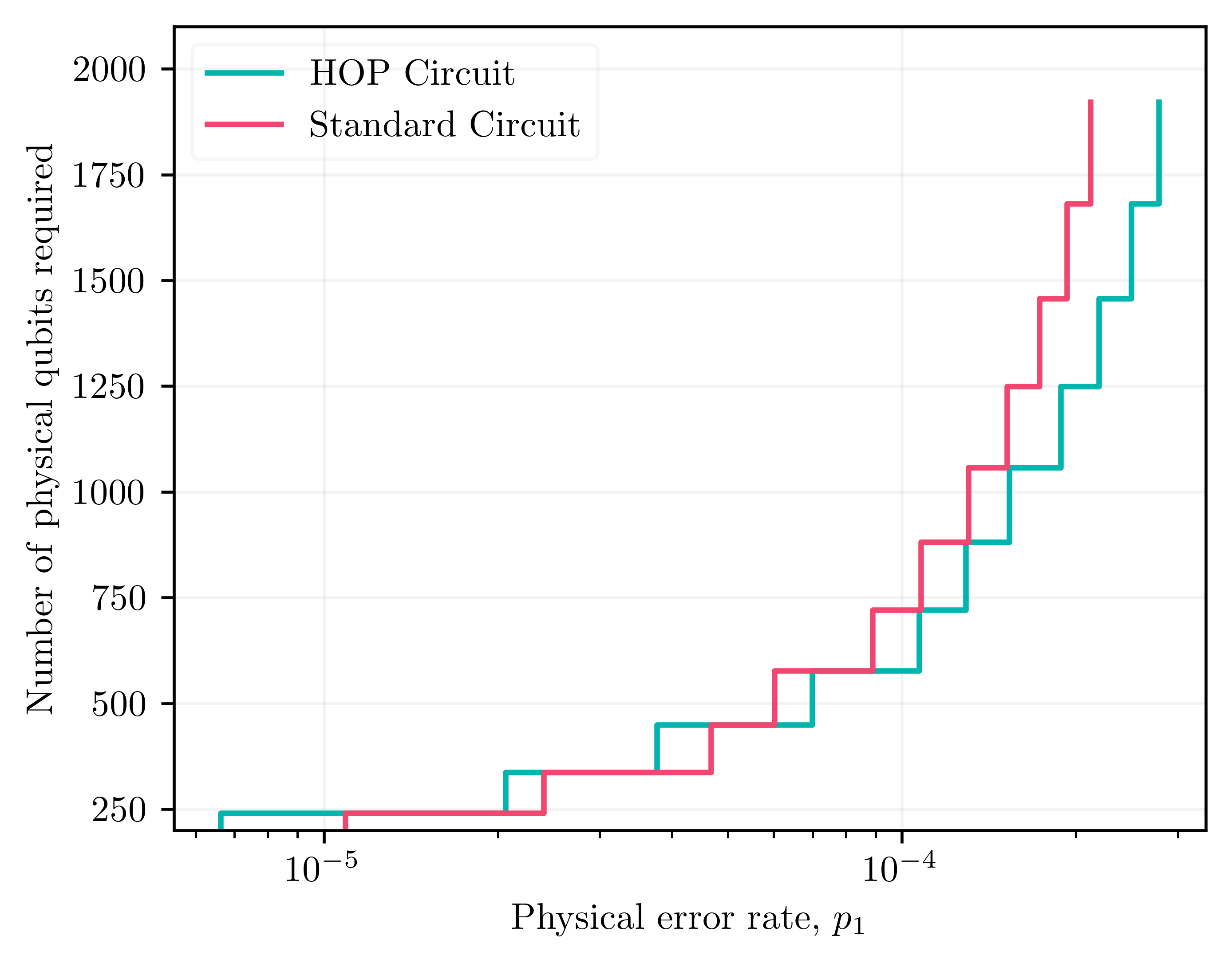}
\caption{\label{fig:QubRes} Comparison of the total number of physical qubits required in order to implement a logical qubit with logical error rate $p_L=10^{-10}$ using either scheme at various physical error rates $p_1$.}
\end{figure}

\begin{figure}
\centering
\includegraphics[width=0.8\textwidth]{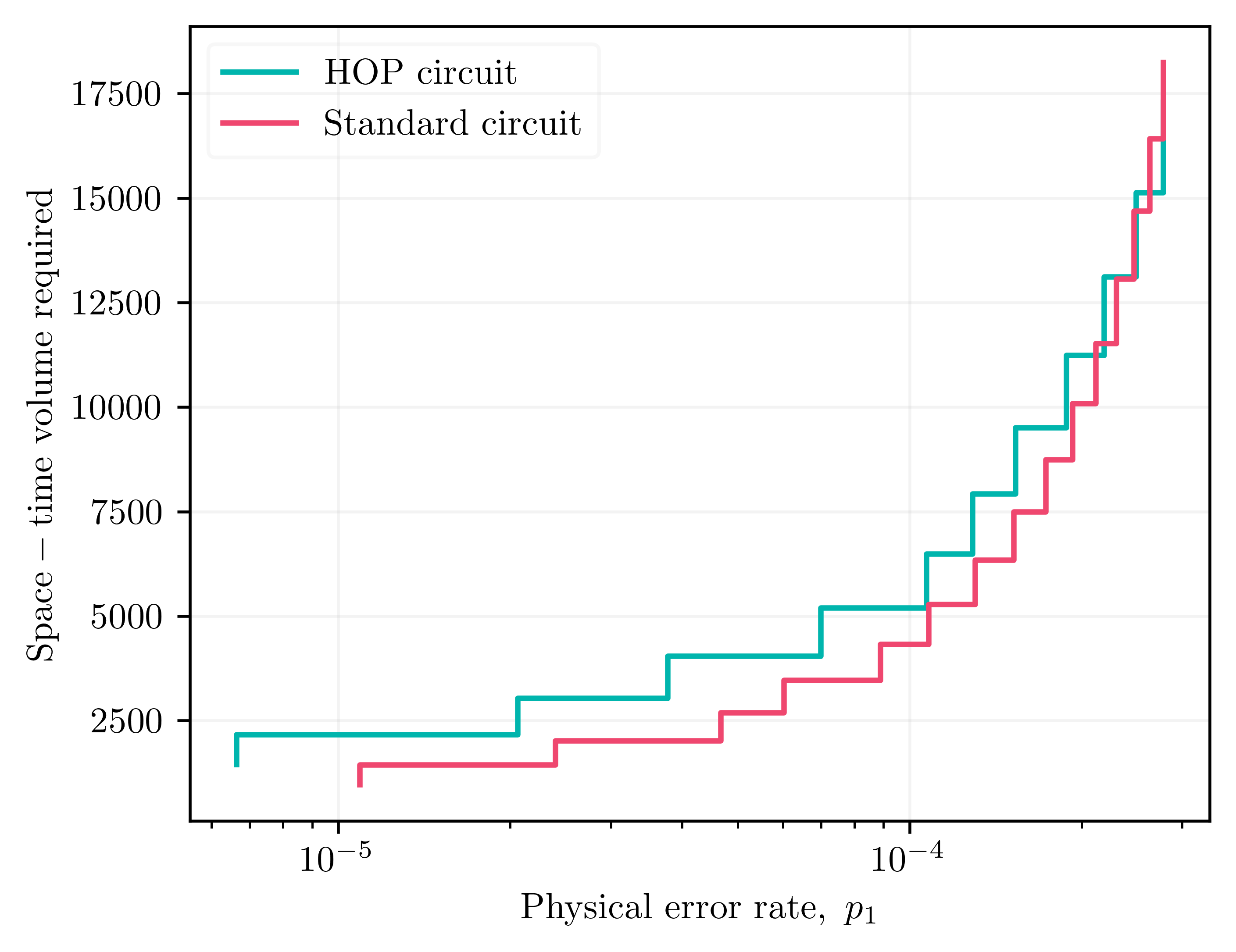}
\caption{\label{fig:STRes} Comparison of the overall space-time volume required in order to implement a single error correction cycle of a logical qubit with logical error rate $p_L=10^{-10}$ using either scheme at various physical error rates $p_1$.}
\end{figure}

We compare estimates of two resources: physical qubit count required to implement a surface code of a distance high enough to achieve the target error rate, as well as the space-time volume (the product of physical qubit count and time steps required to complete a logical error correction cycle). In Figure \ref{fig:QubRes} we see that at physical error rates closer to threshold, we can use a HOP surface code of smaller distance to achieve our target error rate than would be required using the standard surface code, as we would expect based on the fact that the HOP scheme yields a higher threshold. In terms of overall space-time volume, we see from Figure \ref{fig:STRes} that the two schemes are relatively evenly matched at higher physical error rates, but the standard scheme has an advantage at lower physical error rates. This is due to a single round of parity check measurements requiring only 6 time steps in the standard scheme while the HOP scheme requires 9.

\section{Conclusions}
We have presented simulated results of a surface code generated using a parity check operation defined by a strong, five-qubit dispersive $ZZ$ interaction between data qubits and their corresponding stabilizer qubits, called the Hardware Optimized Parity (HOP) gate. We found a circuit-level error threshold for the HOP gate of approximately $1.25 \times 10^{-3}$ using conservative estimates for the code construction. For realistic HOP gate parameters of $\zeta_0/2\pi=5~\text{MHz}$, this corresponds to a threshold for decoherence rates as $T_2=80~\mu$s, which is readily exceeded with state-of-art transmon devices~\citep{Place_2021,Siddiqi_2021}. Potential advantages are reinforced by the space-time resource estimates near the threshold, though we note a resource advantage for the standard circuit at lower physical error rates.

We believe additional improvements are likely attainable for this design. We point out that with the schedule A/B/C/D, it is possible to actuate simultaneous stabilizer measurements of the same type \{A,B\} and \{C,D\}, since the Hamiltonian for these HOP gates are mutually commuting. Moreover, an error in one qubit during the HOP gate causes, at worst, a phase flip on one of the gate targets. However, phase errors commute with the HOP Hamiltonian, meaning errors would remain local during simultaneous HOP gates, even though data qubits undergo multiple HOP gates in parallel. Investigating such optimizations is left for future work.

Further one can imagine extending HOP gate type calibration and analysis to larger subsets of the code. For example, one could design and calibrate a 13-qubit operator to perform all the ABCD steps of Figure 5b in one operator. This would push the limits of today's calibration, but could be feasible in the future

Another intriguing possibility is operating the HOP based surface code at a static flux bias $\zeta(t)=\zeta_0$. Removing active controls for entangling gates would address a major challenge towards managing the thermal heat load of large-scale superconducting processors---approximately halving the dissipation at cryogenic temperatures. First, for systems running HOP gates exclusively for entanglement, all DC current bias can be delivered over twisted pair (TwP) superconducting wire (for instance, niobium-titanium alloys) instead of coaxial signal delivery, which is otherwise standard for high-bandwidth two-qubit gate signals~\citep{DiCarlo_2009}. This rewiring, alone, would amount to a hundredfold reduction in passive heat load at the coldest stages of the dilution refrigerator for two-qubit signals~\citep{Krinner_2019,Hollister_2021}, eliminating approximately 40$\%$ of the total estimated thermal budget~\citep{Krinner_2019}. Moreover, another $10\%$ of total thermal budget would be saved by eliminating the active heat load from fast two-qubit gate pulses~\citep{Krinner_2019}. Maintaining an equal thermal footprint while doubling the number of physical qubits could prove to be a leading motivation for HOP gates. Note that Hadamard gates in this architecture must be done unconditionally on the state of other qubits in the presence of the $ZZ$ interaction, which has been demonstrated with high-fidelity for analogous experiments with strong-dispersive interactions in cavity-transmon systems using either large bandwidth control~\citep{Kirchmair2013} or optimal control~\citep{Heeres2015}. Exploring these ideas with near-term hardware will be an essential step in understanding the implications.

\section{Acknowledgements}

This work was jointly sponsored by Rigetti Computing and Goldman Sachs. We thank Stefano Poletto and Angela Chen for their useful input on hardware parameters that helped with calculations of next-nearest neighbor interactions. We thank Craig Gidney and Oscar Higgott for assistance in troubleshooting the use of Stim and PyMatching respectively. We thank Michael Beverland for valuable feedback on early simulation results. Results in this work was made possible using a combination of open-source software utilities as cited in the main text.

\appendix
\section{Parameters for nearest neighbor  and next-nearest neighbor  ZZ couplings for 5-qubit system}\label{app:nnn}
The nearest neighbor  and next-nearest neighbor ZZ couplings are calculated by diagonalizing the full five-qubit Hamiltonian given by Eq. \eqref{eq:q5}. We used realistic parameters \cite{Sete2021} for asymmetric tunable coupler and parameters of the qubits  for numerical diagonalization to computed ZZ couplings.

The nearest neighbor ZZ couplings for between the stabilizer and data qubits are
\begin{align}
\zeta_{01} &= \tilde\omega_{|10100\rangle}+\tilde\omega_{|00000\rangle}-\tilde\omega_{|10000\rangle}-\tilde\omega_{|00100\rangle}, \\ 
\zeta_{02} &= \tilde\omega_{|01100\rangle}+\tilde\omega_{|00000\rangle}-\tilde\omega_{|01000\rangle}-\tilde\omega_{|00100\rangle}, \\
\zeta_{03} &= \tilde\omega_{|00110\rangle}+\tilde\omega_{|00000\rangle}-\tilde\omega_{|00100\rangle}-\tilde\omega_{|00010\rangle}, \\
\zeta_{04} &= \tilde\omega_{|00101\rangle}+\tilde\omega_{|00000\rangle}-\tilde\omega_{|00100\rangle}-\tilde\omega_{|00001\rangle}
\end{align}
and the next-nearest neighbor couplings between data qubits are given by
\begin{align}
\zeta_{12} &= \tilde\omega_{|11000\rangle}+\tilde\omega_{|00000\rangle}-\tilde\omega_{|10000\rangle}-\tilde\omega_{|01000\rangle},\\
\zeta_{13} &= \tilde\omega_{|10010\rangle}+\tilde\omega_{|00000\rangle}-\tilde\omega_{|10000\rangle}-\tilde\omega_{|00010\rangle},\\
\zeta_{14} & = \tilde\omega_{|10001\rangle}+\tilde\omega_{|00000\rangle}-\tilde\omega_{|10000\rangle}-\tilde\omega_{|00001\rangle},\\
\zeta_{23} & = \tilde\omega_{|01010\rangle}+\tilde\omega_{|00000\rangle}-\tilde\omega_{|01000\rangle}-\tilde\omega_{|00010\rangle}, \\
\zeta_{24} & = \tilde\omega_{|01001\rangle}+\tilde\omega_{|00000\rangle}-\tilde\omega_{|01000\rangle}-\tilde\tilde\omega_{|00001\rangle},\\
\zeta_{34} & = \tilde\omega_{|00011\rangle}+\tilde\omega_{|00000\rangle}-\tilde\omega_{|00010\rangle}-\tilde\omega_{|00001\rangle}. 
\end{align}

\begin{table}[ht]
    \centering
    \begin{tabular}{c|c}
    \hline
    \hline
         parameter & value  \\
         \hline
        $g_{0j}/2\pi$ & -8 MHz\\
        $g_{0c_1}/2\pi$ & 130 MHz\\
        $g_{0c_2}/2\pi$ & 132 MHz\\
        $g_{0c_3}/2\pi$ & -112 MHz\\
        $g_{0c_4}/2\pi$ & -130 MHz\\
        $g_{1c_1}/2\pi$ & -149.5 MHz\\
        $g_{2c_2}/2\pi$ & -134.5 MHz\\
        $g_{3c_3}/2\pi$ & 134.5 MHz\\
        $g_{4c4}/2\pi$ & 143.0 MHz\\
        \hline
    \end{tabular}
    \caption{Stabilizer-data qubit couplings and qubit-tunable coupler couplings used in the numerical diagonalization of the 5-qubit Hamiltonian \eqref{eq:q5} for computing the ZZ couplings.}
    \label{tab:couplings}
\end{table}

\begin{table}[t!]
    \centering
    \begin{tabular}{cccccccccc}
    \hline
    \hline
       parameters  & q$_0$ & q$_1$ & q$_2$ & q$_3$ & q$_4$ & c$_1$& c$_2$ & c$_3$ & c$_4$ \\
       \hline
      $\omega_{\rm max}/2\pi(\rm{GHz})$  & 4.0 &3.85 & 3.80 & 4.15 & 4.20 & 6.7  & 6.7 & 6.7 & 6.7\\
    $\omega_{\rm min}/2\pi (\rm{GHz})$  & - &- & -&- & - & 4.58  &4.75 & 4.9 & 4.9\\
    $\eta_{\rm max}/2\pi (\rm{MHz})$  & 185 &221 & 190&216 & 220 & 200  &200 & 200 & 200\\
       \hline
    \end{tabular}
    \caption{Frequencies and anharmonicities of the qubits and tunable couplers used in the numerical diagonalization of the 5-qubit Hamiltonian \eqref{eq:q5} for computing the ZZ couplings.}
    \label{tab:freqs}
\end{table}
\section{Approximating a Correlated Pauli Channel}

As discussed in Sec.~\ref{sec:error_model}, the 5Q and 3Q interactions are represented in our circuit-level error model via perfect entangling gates followed by a correlated 5Q or 3Q Pauli channel:
\begin{equation}
        \mathcal{E}_{5/3}(\rho)=\sum_{A\in\mathcal{P}_{5/3}}p_A A\rho A^\dagger,
\end{equation}
where $p_A$ is the probability that pauli $A\in\mathcal{P}_{5/3}$ occurs. Recall that the characteristic parameter $p$ of the channel describes the strength of the noise in the system, and that $p> p_A$ for any $A$.

Stim, however, does not support the use of correlated Pauli channels with more than two outcomes beyond two qubits. In this appendix, we discuss and justify how we approximated our intended noise model using the channels that Stim natively permits. Stim allows the use of a channel that applies an arbitrary multi-qubit Pauli operator with probability $p$, and otherwise applies the identity, with probability $1-p$:
\begin{equation}
        \mathcal{E}_{A,p}(\rho)=(1-p)\rho + pA\rho A^\dagger,~A\in\mathcal{P}_n
\end{equation}
We will refer to such channels as single-outcome Pauli channels.

A simple way of approximating $\mathcal{E}_{5/3}$ is to replace it with a channel $\mathcal{E}'_{5/3}$ which sequentially applies a separate single-outcome Pauli channel $\mathcal{E}_{A,p_A}$ for each non-trivial Pauli operator $A\in\mathcal{P}_{5/3}$ with its corresponding probability $p_A$. Our desired channel $\mathcal{E}_{5/3}$ applies Pauli $A$ with probability $p_A$, and by comparison, our approximate channel $\mathcal{E}'_{5/3}$ will apply Pauli $A$ with probability
\begin{equation}
        p'_{A}=p_A + C'_{A,p^2}+C'_{A,p^3}+...+C'_{A,p^{4^{5/3}}},
\end{equation}
where $C'_{A,p^k}$ is the sum of all probabilities that exactly $k$ of the single-outcome channels of $\mathcal{E}'_{5/3}$ apply non-trivial Paulis whose product gives the Pauli $A$. Because $p> p_A$, we see that
\begin{equation}
        p'_{A}=p_A + O(p^2).
\end{equation}
This tells us that $\mathcal{E}'_{5/3}$ approximates $\mathcal{E}_{5/3}$ to first order in $p$, and could be a reasonable approximation for small values of $p$.

We can improve this approximation to order $O(p^2)$ with a reasonable computational overhead by computing $C'_{A,p^2}$ for each $A\in\mathcal{P}_{5/3}$ and shifting the probabilities for each of the single-outcome Pauli channels to $p_A - C'_{A,p^2}$. We refer to the channel built with this procedure as $\mathcal{E}''_{5/3}$.

Lemma: The channel $\mathcal{E}''_{5/3}$ applies pauli $A$ with probability $p_A+O(p^3)$.

Proof: Under the shift $p_A\to p_A - C'_{A,p^2}$, we have that the probability of applying $A$ is
\begin{equation}
        p''_{A}=p_A - C'_{A,p^2}+C''_{A,p^2}+C''_{A,p^3}+...+C''_{A,p^{4^{5/3}}},
\end{equation}
where the terms $C''_{A,p^k}$ are defined appropriately with the shifted probabilities. Examining $C''_{A,p^2}$, we have that
\begin{equation}
        C''_{A,p^2} = \sum_{B,C~s.t.~BC\sim A}(p_B-C'_{B,p^2})(p_C-C'_{C,p^2})
        \\
        =\sum_{B,C~s.t.~BC\sim A}p_Bp_C + O(p^3)
        \\
        =C'_{A,p^2} + O(p^3),
\end{equation}
so that
\begin{equation}
        p''_{A}=p_A - C'_{A,p^2}+C'_{A,p^2}+O(p^3)
        \\
        =p_A + O(p^3)
\end{equation}
as desired.

We see, then, that allowing Stim to implement the channel $\mathcal{E}''_{5/3}$ instead of $\mathcal{E}_{5/3}$ should not have a meaningful impact on simulations at low $p$.

\section{Linear Fit Data for Error Correction Simulations}
\begin{figure}
\centering
\includegraphics[width=0.8\textwidth]{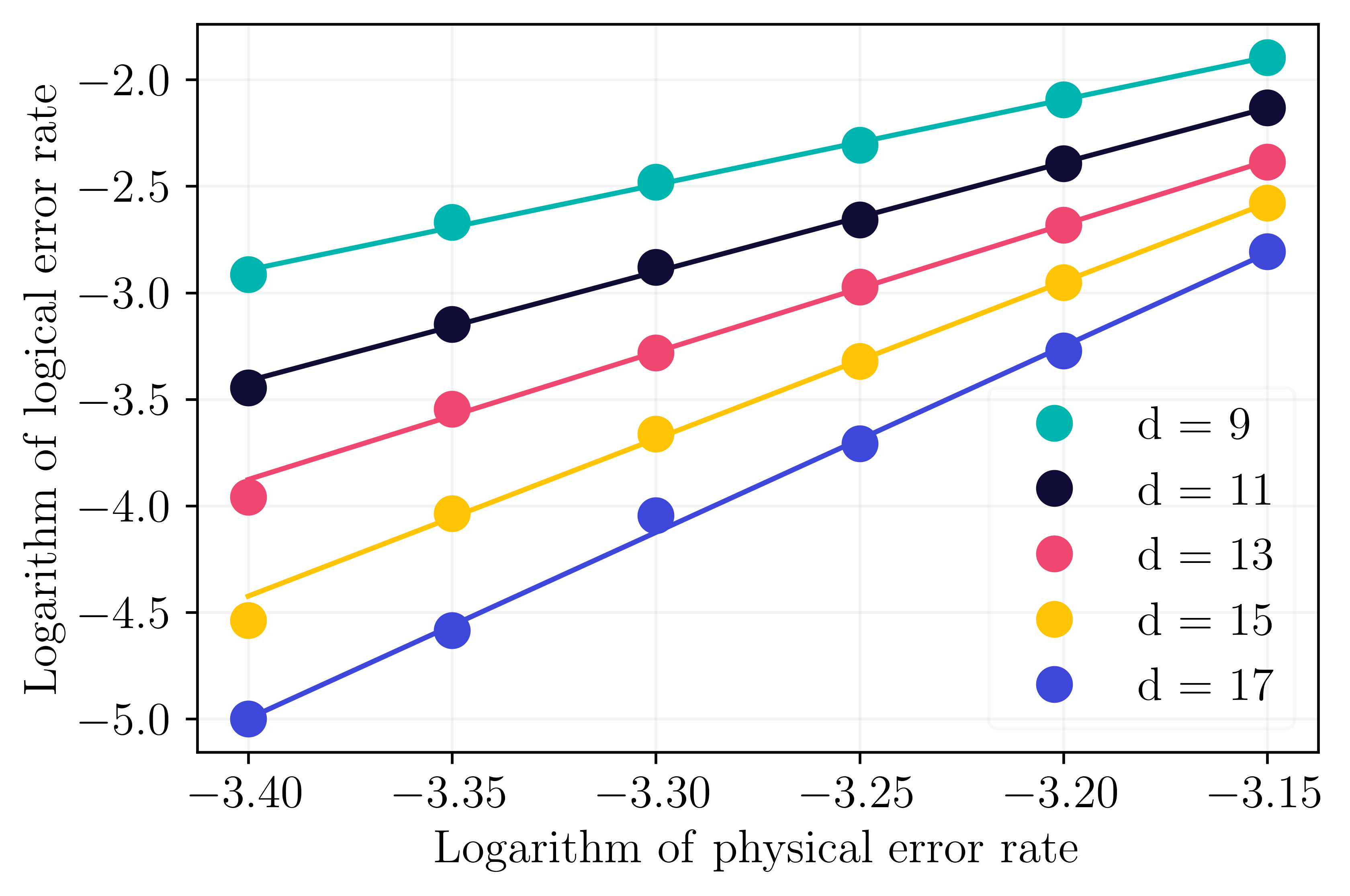}
\caption{\label{fig:HOPFits} Linear fits of lower $p_1$ simulation data for the HOP scheme.}
\end{figure}

\begin{figure}
\centering
\includegraphics[width=0.8\textwidth]{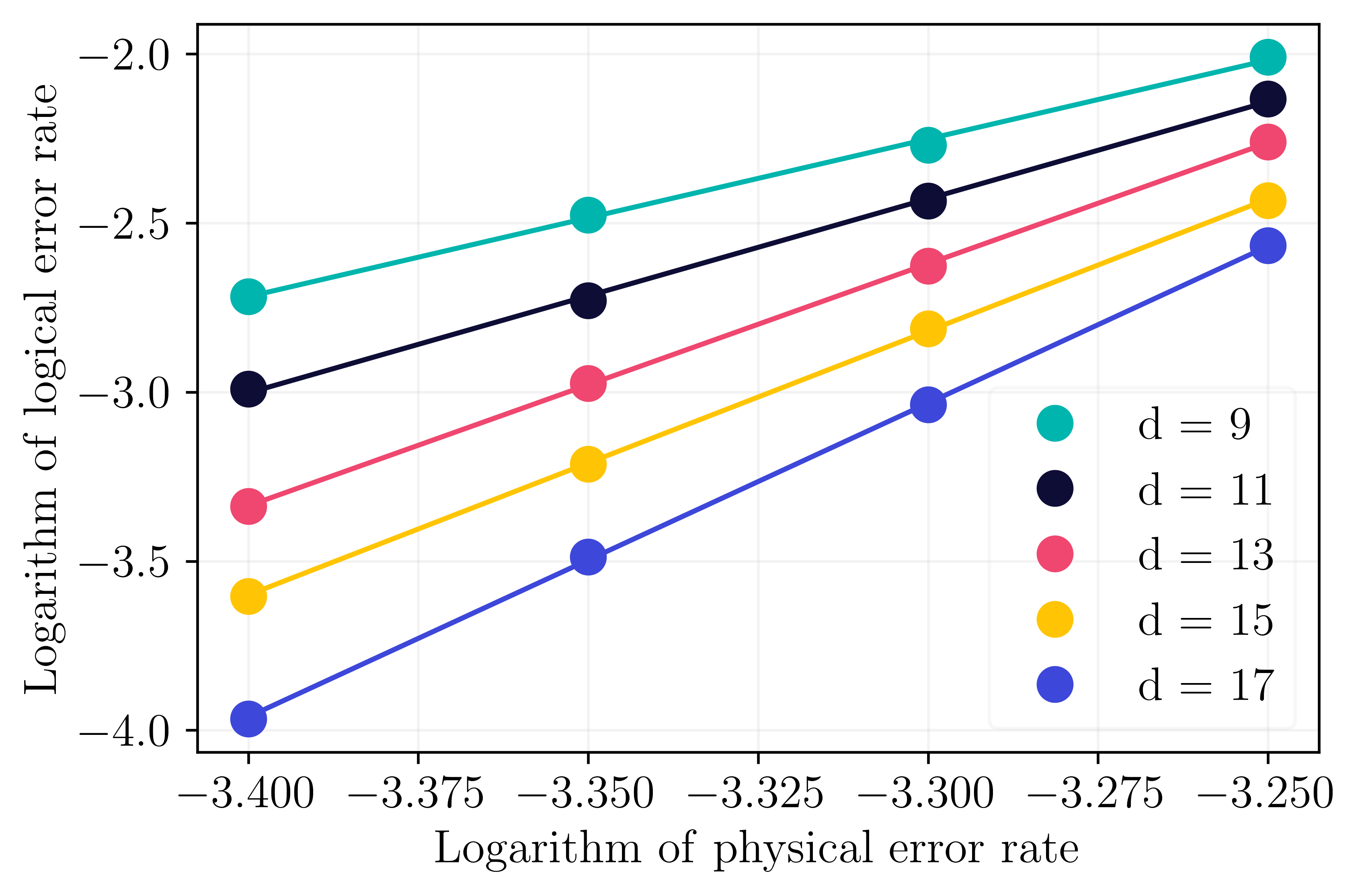}
\caption{\label{fig:StdFits} Linear fits of lower $p_1$ simulation data for the standard scheme.}
\end{figure}

In order to estimate the code distances required in order to achieve target physical error rates outside of the simulated regime, we perform linear fits in logarithmic space to extract the logical error rate's functional dependence on the code distance $d$ and physical error rate $p_1$. We use the simple heuristic \cite{Fowler_2012}
\begin{equation}
        p_L(p_1,d)=c(d)p_1^{m(d)},
\end{equation}
where we expect $m(d)$ to be a linear function of $d$ and $c(d)$ to be an exponential function of $d$. Tables \ref{tab:HOPfits} and \ref{tab:Stdfits} list the fit parameters obtained for the HOP and standard schemes respectively, while Figures \ref{fig:HOPFits} and \ref{fig:StdFits} show the lines fitted to the data points.

\begin{table}[h!]
  \begin{center}
    \caption{Table describing fit parameters obtained from linear regression for the HOP scheme.}
    \label{tab:HOPfits}
    \begin{tabular}{l|c|r} 
      $d$ & $\log c(d)$ & $m(d)$ \\
      \hline
      9 & 10.69 & 3.99\\
      11 & 14.00 & 5.12\\
      13 & 16.46 & 5.98\\
      15 & 20.65 & 7.37\\
      17 & 24.70 & 8.73\\
      19 & 26.62 & 9.41\\
      21 & 28.60 & 10.11\\
    \end{tabular}
  \end{center}
\end{table}

\begin{table}[h!]
  \begin{center}
    \caption{Table describing fit parameters obtained from linear regression for the standard scheme.}
    \label{tab:Stdfits}
    \begin{tabular}{l|c|r} 
      $d$ & $\log c(d)$ & $m(d)$ \\
      \hline
      9 & 13.13 & 4.66\\
      11 & 16.51 & 5.74\\
      13 & 21.00 & 7.16\\
      15 & 22.93 & 7.80\\
      17 & 27.58 & 9.27\\
    \end{tabular}
  \end{center}
\end{table}

By performing linear regression on the data points obtained for $m(d)$ and $\log c(d)$, we can estimate $m(d)$ and $\log c(d)$ for $d$ larger than we were able to practically simulate, and use these extrapolated values to produce Figures \ref{fig:QubRes} and \ref{fig:STRes} in the main text. We found the lines of best fit for $m(d)$ for each scheme to be
\begin{equation}
        m_{HOP}(d)=0.53 d -0.71
\end{equation}
and
\begin{equation}
        m_{Std}(d)=0.56 d -0.406
\end{equation}
whereas the lines of best fit for $\log c(d)$ were found to be
\begin{equation}
        \log c_{HOP}(d)=1.56 d -3.11
\end{equation}
and
\begin{equation}
        \log c_{Std}(d)=1.77 d -2.73.
\end{equation}

\bibliographystyle{unsrtnat}
\bibliography{main}

\end{document}